\documentclass[pra,twocolumn,tightenlines,nofootinbib]{revtex4}
\usepackage{bm,dcolumn,amsmath,graphicx,amsfonts,amssymb}

\newcommand{\tref}[1]{Table~\ref{#1}}

\begin{document}
\title{Nobelium energy levels and hyperfine structure constants}
\author{S.~G.~Porsev$^{1,2}$}
\author{M.~S.~Safronova$^{1,3}$}
\author{U.~I.~Safronova$^{4}$}
\author{V.~A.~Dzuba$^{5}$}
\author{V.~V.~Flambaum$^{5,6}$}
\affiliation{
$^1$Department of Physics and Astronomy, University of Delaware,  Newark, Delaware 19716, USA\\
$^2$Petersburg Nuclear Physics Institute of NRC ``Kurchatov Institute'', Gatchina, Leningrad district, 188300, Russia\\
$^3$Joint Quantum Institute, NIST and the University of Maryland, College Park, Maryland 20742, USA\\
$^4$Physics Department, University of Nevada, Reno, Nevada 89557, USA \\
$^5$School of Physics, University of New South Wales, Sydney 2052, Australia \\
$^6$Helmholtz Institute Mainz, Johannes Gutenberg University, 55099 Mainz, Germany}
\date{ \today }

\begin{abstract}
Advances in laser spectroscopy of superheavy ($Z>100$) elements enabled determination
of the  nuclear moments of the heaviest nuclei, which requires  high-precision atomic calculations
of the relevant hyperfine structure (HFS) constants. Here, we calculated the HFS constants and energy levels for a number of
nobelium ($Z=102$) states using the hybrid approach, combining linearized coupled-cluster and configuration interaction methods.
We also carried out an extensive study of the No energies using 16-electron configuration interaction method to determine the position of the
$5f^{13} 7s^2 6d$ and $5f^{13}7s^2 7p$  levels with a hole in the $5f$ shell to evaluate their potential effect on the hyperfine structure calculations of the low-lying $5f^{14}7s 6d$ and $5f^{14}7s7p$ levels. We find that unlike the case of Yb, the mixing of the low-lying levels with filled
and unfilled $f$ shell is small and does not significantly influence their properties. The resulting HFS constants for the $5f^{14}7s7p\,\,^1\!P_1^o$ level, combined with laser-spectroscopy measurement,  were used to extract nobelium nuclear properties
[S. Raeder \textit{et al.}, Phys. Rev. Lett. {\bf 120}, 232503 (2018)].

\end{abstract}

\maketitle
\section{Introduction}
\label{Intro}

A study of superheavy element properties is a very important and challenging task
that requires a development of new experimental and theoretical methods.
Only very limited experimental information about properties of the superheavy elements is available
due to their low production rate of only a few atoms per second at most. These radioactive elements must be studied immediately following their production in nuclear fusion reactions. Experimental data regarding electronic configurations and ionization potentials of these elements are very scarce and theoretical calculations are required to obtain this information. The nuclear moments of the heaviest nuclei could only be inferred from nuclear spectroscopy requiring model assumptions until recent laser spectroscopy advances \cite{LaaLauBac16,RaeAckBac18}.

Atomic spectra of different isotopes of superheavy elements can be used to obtain information on the nuclear spin, nuclear
moments, and changes in nuclear mean-charge radii between isotopes allowing direct probes of nuclear properties.
Atom-at-a-time laser resonance ionization spectroscopy of  nobelium was reported in \cite{LaaLauBac16}, in which the
$7s^2\,\, ^1\!S_0 - 7s7p\,\, ^1\!P_1^o$ transition was identified.
Further laser spectroscopy studies of this No transition were carried out in \cite{RaeAckBac18} including the measurement of the hyperfine splitting of $^{253}$No and the isotope shifts for $^{252,253,254}$No.
Combining these measurements with the state-of-the-art atomic calculations allowed an extraction of the nuclear properties such as the nuclear magnetic dipole and electric quadrupole moments of No and change of the nuclear radius between 252, 253, and 254 isotopes~\cite{RaeAckBac18}.
In this work we describe these calculations in detail and present recommended values for a number of No hyperfine structure (HFS) constants for a future improvement of the nuclear properties determination.

A problem that occurs in the No calculation is the electronic structure of the low-lying levels. The configurations with two electrons
above the closed $5f$ shell, such as $5f^{14}7s 6d$ and $5f^{14}7s7p$, can be treated with most accurate methods of calculation,
such as a combination of the configuration interaction (CI)~\cite{KotTup87} with many-body perturbation theory (the CI+MBPT method) or with a coupled-cluster approach (the CI+all-order method) (see~\cite{DzuFlaKoz96,SafKozJoh09,Dzu14} for more details).
However, these methods cannot reproduce the energy levels belonging to the $5f^{13}7s^2 6d$ and $5f^{13}7s^2 7p$ configurations, which have a hole in the $5f$ shell, and, hence, a mixing of these configurations with $5f^{14}7s7p$ and $5f^{14}7s 6d$.
Therefore, if such states appear low in the spectra, the CI+MBPT or CI+all-order methods may not be reliable. On the other hand, if these states appear to be high in the spectra they will not affect the properties of the low-lying states with filled $5f$ shell.

Nobelium is a chemical homolog of Yb and it is known~\cite{RalKraRea11} that the Yb energy levels  with unfilled $4f$ shell already appear at a level of 23000 cm$^{-1}$. It leads to a significant mixing of these states with the states with filled $4f$ shell, particularly strongly affecting the properties of the $4f^{14}6s6p\,\,^1\!P_1^o$ level and resulting in a poor accuracy of theoretical HFS constants for this state~\cite{PorRakKoz99,DzuDer10}.
To check whether this is also the case for nobelium, whose main configuration is $5f^{14}\, 7s^2$, we consider it as a system with 16 valence electrons
and perform calculations of the low-lying energy levels in the framework of (i) the conventional CI method and
(ii) recently developed method based on a CI technique, where excitations of the valence electrons to high-energy states are treated perturbatively
(the CIPT method)~\cite{DzuBerHar17}. It allows us to determine the position of the states with filled and unfilled $5f$ shell relative to each other. Both these methods do not take into account the core-valence correlations and, hence, are not expected to be as accurate as the CI-all-order method for the divalent states. But they can deal with many-valence atoms and ions giving a reasonable calculation accuracy.

Our analysis shows that an interaction of the states with filled and unfilled $5f$ shell is practically negligible, what allows us to
consider No atom as a divalent system and apply the CI+all-order method~\cite{SafKozJoh09} (combining CI with the linearized single-double coupled-cluster (LCCSD) method) for calculating the HFS constants of the low-lying states.
We find that No case is very similar to Hg, where core-excited states appear much higher in the spectrum, not significantly affecting the accuracy of the $6s6p\,\,^1\!P_1^o$ HFS constants. We start with a description of these energy studies and then consider the HFS constants.
\section{Methods of calculation}
\label{Methods}
Here we consider No as a system with 16 valence electrons and perform calculations in the framework of the CI method.
We start from a solution of the Dirac-Fock equations and carry out the initial self-consistency procedure for the [$1s^2,...,5f^{14}\, 7s^2$]
configuration. To optimize the calculations for a particular problem described above, we construct the orbitals for specific configurations. The $7p$ orbitals were constructed for the $5f^{14} 7s7p$ configuration, i.e., freezing all orbitals and moving an electron from $7s$ to $7p$ shell. The $6d$ orbitals were constructed for the $5f^{13}\, 7s^2\, 6d$ configuration.
The virtual orbitals were constructed as described in~\cite{Bog91,KozPorFla96}. In total, the basis set included orbitals up to $9s$, $9p$, $8d$, $8f$, and $7g$. The size of a configuration space grows very rapidly with increasing the basis set. The basis used by us makes the calculation manageable while still allowing to perform convergence tests to ensure the validity of the results. The configuration space was formed by allowing single and double excitations for the even-parity states from the configurations $5f^{14} 7s^2$ and $5f^{14} 7s6d$ and for the odd-parity states from the configurations $5f^{14} 7s7p$ and $5f^{13} 7s^2 6d$.

To verify a convergence of the CI method, we calculated the low-lying energy levels for three cases: including the single and double excitations to the shells $7s, 7p, 6d, 6f$, and $5g$ (we designate it as [$7sp6df5g$]) and including the single and double excitations to [$8sp7df6g$] and
[$9sp8df7g$]. In the last case the configuration space consisted of $2\,460\,000$ determinants for the even-parity states
and $3\,000\,000$ determinants for the odd-parity states presenting already a significant computational challenge.

The results of calculation of the energies, using the three CI spaces described above, are given in~\tref{No_E}. Where available, we compare our results with those obtained in Ref.~\cite{DzuSafSaf14} in the framework of the CI+all-order method, where No was treated as a divalent atom, thus only allowing to obtain results for the $5f^{14}7snl$ configurations. We refer a reader to Ref.~\cite{DzuSafSaf14} for the description of the CI+all-order method and its application to the calculation of the No energy levels.

\begin{table}
\caption{The energy levels of the low-lying excited states of No counted
from the ground state (in cm$^{-1}$). The columns [$7sp6df5g$],
[$8sp7df6g$], and [$9sp8df7g$] give results obtained using different
sets of the configurations described in the text.
The results obtained in Ref.~\cite{DzuSafSaf14} by the CI+all-order method are given in the column labeled ``CI+All''.}
\label{No_E}
\begin{ruledtabular}
\begin{tabular}{llcccc}
 Config.      &     Term            &[$7sp6df5g$]
                                              &[$8sp7df6g$]
                                                         &[$9sp8df7g$]
                                                                    & CI+All \\
\hline \\
$5f^{14}7s^2$    & $^1\!S_0$       &    0    &    0     &    0     &    0    \\
$5f^{14}7s6d$    & $^3\!D_1$       &  35287  &  30139   &  31003   &  28436  \\
                 & $^3\!D_2$       &  35197  &  30354   &  31223   &  28942  \\
                 & $^3\!D_3$       &  35023  &  30722   &  31608   &  30183  \\
                 & $^1\!D_2$       &  41802  &  37230   &  37980   &  33504  \\[0.1cm]

$5f^{13}7s^2 7p$ & $J=3$           &  59856  &  56410   &  56927   &         \\
                 & $J=4$           &  59959  &  56537   &  57068   &         \\
                 & $J=5$           &  68133  &  64396   &  64911   &        \\
                 & $J=2$           &  68220  &  64463   &  64984   &         \\
                 & $J=3$           &  68571  &  64899   &  65434   &         \\[0.3cm]

$5f^{14}7s 7p$   & $^3\!P_0^o$     &  15321  &  16278   &  16360   &  19567  \\
                 & $^3\!P_1^o$     &  17184  &  18064   &  18138   &  21042  \\
                 & $^3\!P_2^o$     &  21609  &  22508   &  22536   &  26113  \\
                 & $^1\!P_1^o$     &  30365  &  30173   &  30237   &  30203  \\[0.1cm]

$5f^{13}7s^2 6d$ & $J=2$           &  44816  &  45492   &  45720   &         \\
                 & $J=5$           &  49126  &  49494   &  49731   &         \\
                 & $J=3$           &  51741  &  51929   &  52172   &         \\
                 & $J=6$           &  51614  &  52075   &  52415   &         \\
                 & $J=4$           &  53354  &  53426   &  53701   &         \\
                 & $J=2$           &  53438  &  53671   &  54016   &         \\
                 & $J=1$           &  55319  &  55357   &  55695   &         \\
                 & $J=4$           &  56120  &  56216   &  56597   &         \\
                 & $J=3$           &  56608  &  56577   &  56958   &
\end{tabular}
\end{ruledtabular}
\end{table}

The CI results are within 10-20\% of those obtained in Ref.~\cite{DzuSafSaf14}, demonstrating sufficient accuracy of our CI approximation.
We find a large energy separation between the states with filled and unfilled $5f$ shell, unlike the case of Yb. For comparison, the energy difference
between the $4f^{14}6s6p\,\,^1\!P_1^o$ and $4f^{13} 6s^2 5d,\,J=1$ states in Yb is only 3800 cm$^{-1}$. For No
the energy difference between the similar terms $5f^{14}7s7p\,^1\!P_1^o$ and $5f^{13} 7s^2 6d, \,J=1$ is found to be $\sim 25000$ cm$^{-1}$.

For a greater confidence we also performed the energy level calculation employing the CIPT method. In contrast with the conventional CI method a full diagonalization of the energy matrix is not needed in this approach and much longer basis set can be used. It can work with sixteen valence electrons including configurations with filled and unfilled $5f$ shell into the CI matrix. This method is very useful since it can deal with complicated elements,
for which the CI+all-order method is not applicable and the CI method is impracticable because the energy matrix is huge.
The energy interval between the $5f^{14}7s7p\,^1\!P_1^o$ and $5f^{13} 7s^2 6d, \,J=1$ states, found
in the framework of the CIPT method, was in several times larger than between similar terms in Yb, thus, confirming the results obtained
by 16-electron CI.

Based on this consideration we conclude that the states with filled and unfilled $5f$ shell are located sufficiently far from each other. A
mixing between them should be small and is not essential in calculating the properties of the low-lying states with filled $5f$ shell.
Thus, we use in the following the CI+all-order method, as most accurate, for calculation of the HFS constants.
\section{Hyperfine structure constants}
\label{hfs}
\subsection{HFS couplings}
The HFS coupling due to nuclear multipole moments may be represented as a scalar product of
two tensors of rank\textrm{\ k},
\begin{equation*}
H_{\mathrm{hfs}}=\underset{k}{\sum}\left( \mathbf{N}^{(k)}\cdot \mathbf{T}^{(k)}\right) ,
\end{equation*}
where $\mathbf{N}^{(k)}$ and $\mathbf{T}^{\left( k\right) }$ act in the
 nuclear and electronic coordinate space, respectively. Using this expression we
 write the matrix element (ME) of the operator $H_{\mathrm{hfs}}$ as
\begin{eqnarray}
&&\langle \gamma'IJ'FM_{F}|H_{\mathrm{hfs}}|\gamma IJ FM_{F} \rangle = (-1)^{I+J'+F} \nonumber \\
&\times& \underset{k=1}{\sum }\langle I ||N^{(k)}|| I\rangle
\langle \gamma'J' ||T^{(k)}||\gamma J \rangle
\left\{
\begin{tabular}{lll}
$I$ & $I$  & $k$ \\
$J$ & $J'$ & $F$%
\end{tabular}
\right\} .
\label{Hhfs}
\end{eqnarray}
Here $I$ is the nuclear spin, $J$ is the total angular momentum of the electrons,
${\bf F} = {\bf I} + {\bf J}$, and $\gamma$ encapsulates all other electronic quantum numbers.

In the following we restrict ourselves to the first two terms in the sum over
$k$, considering only the interaction of magnetic dipole and
electric quadrupole nuclear moments with the electrons, i.e.,
\begin{equation*}
H_{\mathrm{hfs}}\approx \mathbf{N}^{(1)}\cdot \mathbf{T}^{\left( 1\right) }+%
\mathbf{N}^{(2)}\cdot \mathbf{T}^{\left( 2\right) }.
\end{equation*}

We define $\mathbf{N}^{(1)}$ and $\mathbf{N}^{(2)}$ in a dimensionless
form, expressing them through the nuclear magnetic dipole moment $\boldsymbol \mu$
and nuclear electric quadrupole moment $Q$, respectively, as
\begin{eqnarray*}
\mathbf{N}^{(1)} &=& \boldsymbol{\mu}/\mu_{N}, \\
N_{q}^{(2)} &=&Q_{q}^{(2)}/[1\,\mathrm{b}],
\end{eqnarray*}
where $\mu _{N}$ is the nuclear magneton.
The reduced matrix elements $\langle I |N^{(k)}||I \rangle$ ($k=1,2$) are
\begin{eqnarray*}
\langle I ||N^{(1)}|| I\rangle &=&\sqrt{\frac{(2I+1)(I+1)}{I}}%
\frac{\mu }{\mu _{N}}, \\
\langle I ||N^{(2)}|| I\rangle &=&\frac{1}{2}\sqrt{\frac{%
(2I+3)(2I+1)(I+1)}{I(2I-1)}}\left[ \frac{Q}{1\text{\textrm{b}}}\right] .
\end{eqnarray*}
The operator $T_q^{(k)}$ is the sum of the one-particle operators
\begin{eqnarray*}
T_q^{(k)} = \sum_{i=1}^{N_e} \left( T_q^{(k)} \right)_i,
\end{eqnarray*}
where $N_e$ is the number of the electrons in the atom and the expressions for one-particle electronic tensors $T^{(k)}_i$
are given (in the SI units) by
\begin{eqnarray*}
\left(T_{q}^{(1)}\right)_i &=&-\frac{|e|}{4\pi \varepsilon _{0}}\frac{i\sqrt{2}\left(
\boldsymbol{\alpha}_i \cdot \mathbf{C}_{1q}^{(0)}\left( \mathbf{\hat{r}}_i \right)
\right) }{cr_i^{2}} \cdot \mu _{N}, \\
\left(T_{q}^{(2)}\right)_i &=&-\frac{|e|}{4\pi \varepsilon _{0}}\frac{C_{2q}\left(
\mathbf{\hat{r}}_i \right) }{r_i^{3}} \cdot \left[ 1 \text{\textrm{b}}\right] ,
\end{eqnarray*}
where $\boldsymbol{\alpha}_i$ is the Dirac matrix, $\varepsilon_0$ is the dielectric constant,
$\mathbf{C}_{1q}^{(0)}$ is a normalized spherical harmonic, $C_{2q}$ is a normalized spherical function,
$r_i$ is the radial position of the $i$th electron, and $\mathbf{\hat{r}}_i \equiv {\bf r}_i/r_i$.

The formulas connecting the HFS constants
$A$ and $B$ of an atomic state $\left| J\right\rangle $ with the matrix elements
$\langle \gamma J ||T^{(k)}|| \gamma J \rangle $ of the electronic tensors $\mathbf{T}^{(k)}$ are:
\begin{eqnarray*}
A &=& \frac{g_N}{\sqrt{J(J+1)(2J+1)}} \langle \gamma J ||T^{(1)}|| \gamma J \rangle , \\
B &=&-2\left[ \frac{Q}{1\text{\textrm{b}}}\right] \sqrt{\frac{J\,(2J-1)}{(2J+3)(2J+1)(J+1)}} \\
&\times& \langle \gamma J ||T^{(2)}|| \gamma J \rangle ,
\end{eqnarray*}
where $g_N = \mu/(\mu_N\,I)$.
\subsection{Results and estimate of uncertainties}
 In Ref.~\cite{RaeAckBac18} the nuclear ground-state properties were obtained from laser spectroscopy for the isotopes $^{252,253,254}$No.
 Using these measurements and the calculation of the HFS constants $A$ and $B$ for the $5f^{14}7s7p\,^1\!P_1^o$ state,
 the nuclear magnetic-dipole and electric-quadrupole moments were extracted to be $\mu/\mu_N = -0.527$ and $Q= 5.9\,{\rm b}$, respectively.
Below, we use these values for calculation of the HFS constants for the low-lying states of $^{259}$No ($I=9/2$).

This calculation was performed using the CI+all-order method, introduced in~\cite{SafKozJoh09} and applied to
calculation of nobelium energy levels in Ref.~\cite{DzuSafSaf14}. We determine the $A$ and $B$ HFS constants for the even- and odd-parity low-lying states of $^{259}$No, for the future laser spectroscopy studies of other No transitions. The results are summarized in Table~\ref{No_hfs} where we list results of several computations to demonstrate the size of various contributions and evaluate the uncertainties of the results. The first one is two-particle CI which does not include any core corrections to the wave function. The next stage is a combination of CI and MBPT which includes core-valence correlations in the second order of the perturbation theory.

The next, CI+all-order, results include third and higher-order correlations of the valence electrons with the core. This calculation provides state-of-the-art wave functions, with corrections from the entire core being included, and valence-valence correlations accurately treated in the framework of CI. We note that CI+MBPT includes CI and CI+all-order includes CI+MBPT, so these are listed as total values and not additive corrections. Next, we include corrections to the HFS expectation values beyond the correlation corrections to wave functions, which we refer to as the corrections to the HFS operator.
The  random-phase approximation (RPA)  was taken into account to all orders and given separately in the table in the column labeled ``RPA''. The  core Brueckner, two particle, structural radiation, and normalization corrections were calculated in the second order of MBPT (see Ref.~\cite{DzuKozPor98} for more details). They are grouped together as ``Other''.
\begin{table}[tbp]
\caption{Contributions to the magnetic dipole and electric quadrupole  HFS constants $A$ and $B$ (in MHz).
The CI, CI+MBPT, and CI+all-order values, without any corrections to the HFS operators are listed in the columns labeled ``CI'', ``CI+MBPT'', and
``CI+All'', correspondingly. The RPA corrections to the HFS operator are listed in the column labeled ``RPA''. All other corrections to the HFS operator (core Brueckner, two particle, structure radiation, and normalization) are grouped together in the column labeled ``Other''.
The values in column labeled ``Total'' are obtained as the sum of the values in the ``CI+All'', `` RPA'', and ``Other'' columns.}
\label{No_hfs}%
\begin{ruledtabular}
\begin{tabular}{lcccrrr}
                       &   CI   &CI+MBPT &CI+All &  RPA  & Other & Total  \\
\hline \\ [-0.5pc]
$A(7s6d\;^3\!D_1)$     &   783  &  1054  &   989 &   184 &  -235 &   939  \\[0.1pc]
$A(7s6d\;^3\!D_2)$     &  -417  &  -728  &  -658 &   -26 &    47 &  -637  \\[0.1pc]
$A(7s6d\;^3\!D_3)$     &  -560  &  -777  &  -729 &   -53 &    88 &  -694  \\[0.1pc]
$A(7s6d\;^1\!D_2)$     &   112  &   330  &   277 &    63 &   -74 &   266  \\[0.5pc]

$A(7s7p\;^3\!P_1^o)$   & -1415  & -2289  & -2107 &  -288 &   293 & -2102  \\[0.1pc]
$A(7s7p\;^3\!P_2^o)$   &  -858  & -1219  & -1143 &  -156 &   182 & -1118  \\[0.1pc]
$A(7s7p\;^1\!P_1^o)$   &   437  &   883  &   780 &   102 &  -144 &   739  \\[0.3pc]
$A(7s8p\;^3\!P_1^o)$   & -1835  & -2696  & -2537 &  -336 &   387 & -2486  \\[0.1pc]
$A(7s8p\;^3\!P_2^o)$   & -1029  & -1382  & -1314 &  -174 &   205 & -1283  \\[0.1pc]
$A(7s8p\;^1\!P_1^o)$   &   773  &  1260  &  1172 &   154 &  -189 &  1137  \\[0.5pc]

$B(7s6d\;^3\!D_1)$     &   572  &   982  &  928  &   109 &   624 &  1661  \\[0.1pc]
$B(7s6d\;^3\!D_2)$     &   813  &  1384  & 1316  &   507 &   557 &  2380  \\[0.1pc]
$B(7s6d\;^3\!D_3)$     &  1071  &  1538  & 1510  &  1136 &   933 &  3579  \\[0.1pc]
$B(7s6d\;^1\!D_2)$     &  2062  &  1721  & 1958  &  1281 &   679 &  3919  \\[0.5pc]

$B(7s7p\;^3\!P_1^o)$   & -2522  & -2656  & -2824 & -1029 &  -405 & -4258  \\[0.1pc]
$B(7s7p\;^3\!P_2^o)$   &  2663  &  3069  &  3121 &  1777 &   303 &  5201  \\[0.1pc]
$B(7s7p\;^1\!P_1^o)$   &  1279  &  2270  &  2161 &  1342 &  -490 &  3013  \\[0.3pc]
$B(7s8p\;^3\!P_1^o)$   &  -303  &  -202  &  -231 &   -91 &   -50 &  -373  \\[0.1pc]
$B(7s8p\;^3\!P_2^o)$   &   546  &   650  &   648 &   358 &    53 &  1059  \\[0.1pc]
$B(7s8p\;^1\!P_1^o)$   &   546  &   615  &   621 &   314 &   -49 &   886
\end{tabular}
\end{ruledtabular}
\end{table}

Since the CI calculation for two valence electrons has a negligible uncertainty, the main source of the uncertainties is the core-valence correlations.  Therefore, uncertainties in the values of the HFS constants may be estimated
based on differences between the CI+all-order and CI+MBPT values.
The resulting  uncertainties of the magnetic-dipole constants $A$ for the triplet $^3\!D_J$ and $^3\!P_J^o$ states are 5-10\%,
while the uncertainties of the constants for the singlet $^1\!D_2$ and $^1\!P_1^o$ states are slightly worse, 10-20\%.
The RPA and ``Other'' corrections tend to cancel each other for the constants $A$. We carried out an additional analysis demonstrating
that if the core Brueckner and structural radiation corrections are accounted for in all
orders of the perturbation theory, this cancellation becomes even more pronounced in most cases.

It is more complicated to estimate the uncertainties of the constants $B$ because relative role of different corrections
is larger. The magnitude of the ``RPA'' and/or ``Other'' corrections is comparable with the
``CI+All'' value of the constant in certain cases. In addition, in contrast with the constants $A$, the RPA and ``Other'' corrections have the same sign in majority of cases. Roughly estimating the absolute uncertainty to be equal to the magnitude of the correction ``Other'', we assume
the fractional uncertainties of the $B$ constants to be at the level of 20-25\%.

We also use another method to evaluate the accuracy of the HFS constants. Similar calculations of the magnetic-dipole HFS constants,
using the CI+all-order method, were done for Hg for the lowest-lying odd-parity $^3\!P_1$ and $^1\!P_1$ states and different contributions
were analyzed in Ref.~\cite{PorSafSaf17}. Hg is a good testing case for No due to similar mixing of the core-excited states of the
odd-parity configurations with $J=1$. As illustrated by \tref{No_Hg}, relative contributions to Hg and No HFS
constants are similar, with the only exception of the ``Other'' contribution which is two times larger in No due
to larger core and resulting larger size of the core Brueckner corrections.

In Hg, the CI+all-order value of $A(^1\!P_1^o)$
(with no RPA and other corrections) agrees with experiment to 8\%, while the final number differs from the experimental result by 11\%. This is caused by the cancellations in the values of the various corrections and some inconsistency in their accuracy - RPA is accounted for in all orders while other corrections are calculated in the second order of the perturbation theory. Assuming additional uncertainty in the No values in comparison with Hg due to larger core-Brueckner corrections, we estimate the accuracy of the $A(7s7p\,^3\!P_1^o)$ to be on the order of 15\%.

The accuracy of $A(^3\!P_1^o)$ in Hg is much better, 3.5\%. Therefore, we expect that in No the HFS constant $A(^3\!P_1^o)$ is accurate to about 5-6\%  making it a good case for a benchmark comparison with experiment. Both methods give uncertainty estimates that are in reasonable agreement.
\begin{table*}[t]
\caption{Contributions to the magnetic dipole HFS constants $A$ (in MHz) for the two odd-parity levels of $^{259}$No and
$^{201}$Hg ($I=3/2$ and $\mu/\mu_N = -0.5602$). The CI, CI+MBPT, and CI+all-order values, without any corrections to the HFS operators are listed in the columns labeled ``CI'', ``CI+MBPT'', and ``CI+All'', correspondingly. The relative differences of the CI+MBPT and CI values and the CI+all-order and
CI+MBPT values are listed in \% to illustrate the size of the second-order and higher-order corrections to the wave functions. The RPA
corrections to the HFS operator are listed in the column labeled ``RPA''. The relative size (the ratio of the RPA correction and the total value)
is listed in the next column in \%. All other corrections to the HFS operator are grouped together in the column labeled ``Other''.
Their relative size (the ratio of the ``Other'' correction and the total value) is given in the next column in \%.
The values in column labeled ``Total'' are obtained as (CI+All) + RPA + Other. The experimental values for Hg are given in the last column.}
\label{No_Hg}%
\begin{ruledtabular}
\begin{tabular}{llcccccrcrcrc}
 \multicolumn{1}{c}{Ion} & \multicolumn{1}{c}{State} & \multicolumn{1}{c}{CI} & \multicolumn{1}{c}{CI+MBPT} & \multicolumn{1}{c}{Diff. \%}
&\multicolumn{1}{c}{CI+All} & \multicolumn{1}{c}{Diff. \%} & \multicolumn{1}{r}{RPA} & \multicolumn{1}{c}{RPA \%} & \multicolumn{1}{c}{Other}
&\multicolumn{1}{r}{Other \%} & \multicolumn{1}{r}{Total} & \multicolumn{1}{c}{Experiment} \\
\hline \\ [-0.6pc]
No &$7s7p\,\,^3\!P_1^o$ &  -1415 & -2289  & 38\% & -2107  &  -9\%  & -288  & 14\%  &   293 & -14\% & -2102  &    \\
   &$7s7p\,\,^1\!P_1^o$ &    437 &   883  & 51\% &   780  & -13\%  &  102  & 14\%  &  -144 & -19\% &   739  &    \\ [0.3pc]

Hg &$6s6p\,\,^3\!P_1^o$ &  -3924 &  -5829 & 33\% &  -5499 &  -6\%  & -560  & 10\%  &   408 &  -7\% & -5651  & -5454.569(3)$^{\rm a}$ \\
   &$6s6p\,\,^1\!P_1^o$ &    774 &   1593 & 51\% &   1422 & -12\%  &  153  & 11\%  &  -113 &  -8\% &  1462  &  1316$^{\rm b}$
\end{tabular}
$^{\rm a}$Reference~\cite{Koh61}; $^{\rm b}$Reference~\cite{Lur66}.
\end{ruledtabular}
\end{table*}

 We note that the $A$ and $B$ HFS constants of the $7s7p\,^1\!P_1^o$ state were calculated also by other groups using different
 methods~\cite{RaeAckBac18}. All results are in agreement within their uncertainties.
\section{Conclusion}
We calculated the energy levels of the low-lying even- and odd-parity states in the framework of 16-electron CI method
to demonstrate significant reordering of the No energy levels in comparison with the homology Yb.
In contrast with Yb the No  states with a hole in the $f$ shell are lying sufficiently high;
a possible mixing with the states with filled $f$ shell is small and does not influences significantly on their properties.
As a result, the low-lying divalent No levels can be reliably treated with the CI+all-order method.

We predicted the values for $7s7p$, $7s6d$, and $7s8p$ magnetic-dipole and electric-quadrupole HFS constants using the CI+all-order method, also incorporating the different corrections to the HFS operators. The uncertainties of the recommended values are estimated.
We find that the theoretical accuracy for the $7s7p\,\,^3\!P_1^o$ HFS constant $A$ is expected to be factor of 3 better in comparison with the
$7s7p\,\,^1\!P_1^o$ level making it particularly attractive for future more precise determination of No nuclear properties.

\section*{Acknowledgments}
We are grateful to M. Kozlov for helpful discussions.
This work was supported by U.S. NSF Grant No. PHY-1620687 and the Australian Research Council. M.~S. thanks the School of Physics at UNSW, Sydney, Australia for hospitality and acknowledges support from the Gordon Godfrey Fellowship UNSW program.

\end{document}